# BRICS countries and scientific excellence:

## A bibliometric analysis of most frequently-cited papers

Lutz Bornmann[a], Caroline Wagner[b], and Loet Leydesdorff[c]

First and corresponding author:

[a] Division for Science and Innovation Studies

Administrative Headquarters of the Max Planck Society

Hofgartenstr. 8,

80539 Munich, Germany.

E-mail: bornmann@gv.mpg.de

[b] Battelle Center for Science & Technology Policy,

John Glenn School of Public Affairs, Ohio State University,

Columbus, OH (USA);

E-mail: wagner.911@osu.edu

[c] Amsterdam School of Communication Research (ASCoR),

University of Amsterdam, Kloveniersburgwal 48,

1012 CX Amsterdam, The Netherlands;

E-mail: loet@leydesdorff.net

**Abstract**

The BRICS countries (Brazil, Russia, India, China, and South Africa) are noted for their increasing participation in science and technology. The governments of these countries have been boosting their investments in research and development to become part of the group of nations doing research at a world-class level. This study investigates the development of the BRICS countries in the domain of top-cited papers (top 10% and 1% most frequently cited papers) between 1990 and 2010. To assess the extent to which these countries have become important players on the top level, we compare the BRICS countries with the top-performing countries worldwide. As the analyses of the (annual) growth rates show, with the exception of Russia, the BRICS countries have increased their output in terms of most frequently-cited papers at a higher rate than the top-cited countries worldwide. In a further step of analysis for this study, we generate co-authorship networks among authors of highly cited papers for four time points to view changes in BRICS participation (1995, 2000, 2005, and 2010). Here, the results show that all BRICS countries succeeded in becoming part of this network, whereby the Chinese collaboration activities focus on the USA.

# 1 Introduction

BRIC is the acronym for four major emerging national economies that showed spectacular economic growth during the 1990s: Brazil, Russia, India, and China (O'Neill, 2001). In 2010, South Africa claimed to add an "S" to the end of the acronym, and this nation is now an official member of the BRICS group of influential developing nations (Graceffo, 2011). "The BRICS countries are thought to have the capacity to 'change the world' on account of both the threats and the opportunities they represent from the economic, social and political points of view" (Cassiolato & Lastres, 2011, p. 1). The BRICS have some characteristics in common from a science and technology standpoint, too: They share a meaningful strategic position (in science) on their continents; they contribute significantly to the world's total population of scientists and engineers; they have huge regional disparities in human, economic and scientific development; and they have been investing a lot of money in developing infrastructure for research and development (Cassiolato & Lastres, 2011; Kumar & Asheulova, 2011). They are viewed as models for other developing countries in science and technology (Adams, Pendlebury, & Stembridge, 2013).

Kumar and Asheulova (2011) and Adams, et al. (2013) document the rapid rise in scientific output of the BRICS. In this study, we expand the analysis and investigate bibliometrically the development of the BRICS countries in their contribution to the most prestigious scientific publications by comparing them with other top-performing nations in citation impact and network conductivity.[1] We assess whether the BRICS are advancing in terms of their respective knowledge bases. The focus on the most frequently-cited papers enables us to raise the question about accession of BRICS to an elite structure in the publication system (Bornmann, de Moya-Anegón, & Leydesdorff, 2010; Narin, Stevens, & Whitlow, 1991; Tijssen, Visser, & van Leeuwen, 2002).

---

[1] Leydesdorff, Wagner, and Bornmann (in preparation) use a similar approach: They explore the longitudinal development for the comparison between the EU28, USA, and one of the BRICS countries – China – at the global level, and for the decomposition of the EU28 both in terms of member states and as a network of international co-authorship relations. The study adds the perspective of using the proportions of most frequently cited publications to the raw counts of numbers.

## 2 Methods

In a first step of analysis, we identify the best performing countries worldwide in terms of citation impact. In a second step, we compare the BRICS countries with the best performing countries in terms of their ability to publish highly-cited papers over the last twenty years (1990 to 2010). In a third step, we investigate whether the BRICS countries have become part of the co-authorship network of highly-cited papers. Step 1 is intended only to identify the best-performing countries worldwide which were then included in the analyses of step 2 and 3.

The analyses of steps 2 and 3 are closely related, because one can assume that an important requirement for becoming one of the best performing countries is the integration into the co-authorship network of these countries (see here Bornmann & Marx, 2012). Other assumed requirements on the country level are for example investments in the infrastructure for research and development and the stimulation of exchange of (young) scientists with other (high performing) countries.

### 2.1 Data

The data were drawn in March 2014 from an analytical version of the Web of Science (WoS) at the Max Planck Society (Munich, Germany), which combines the Science Citation Index Expanded (SCI-E), the Social Science Citation Index (SSCI), and the Arts & Humanities Index (A&HCI). This database is compiled and maintained by the Max Planck Digital Library (MPDL, Munich).

The data are composed of a integer counting of papers classified as articles, reviews, or letters in WoS. Every country that appears on a paper is counted as 1, even when it occurs multiple times on the same paper. WoS data can be used for a comparison of BRICS countries and high performing countries since both groups seem to be equally represented in the data base. According to the results of Wagner and Wong (2012) "high quality science from the BRICs appears to be represented at the same level as more advanced countries" (p. 1009). We downloaded the numbers of records (articles, reviews, and letters) and the numbers of most frequently-cited papers from the MPDL database. The download was restricted to those 30 countries with more than 98,000 records between 1990 and 2010. We chose 98,000 papers as the cut-off in order to include South Africa with 98,635 records.

The in-house database provides the capacity to select the most frequently-cited papers in the different citation indices across fields, document types, and publication years. Cross-field and cross-time-period normalizations of citation impact are required to identify the most frequently-cited papers (Schubert & Braun, 1986). For citation impacts to be normalized, a reference set for each paper is needed; this is provided in the MPDL by considering all papers published in the same year by WoS category, and documents of the same type (Leydesdorff & Opthof, 2011).

The published works are parsed in terms of the percentiles of the citation distribution. Percentiles are an alternative to normalization on the basis of central tendency statistics (arithmetic averages of citation counts) (Bornmann, 2013; Bornmann, Leydesdorff, & Mutz, 2013; Leydesdorff, Bornmann, Mutz, & Opthof, 2011) that is less affected by outliers (papers with huge numbers of citations). Percentiles are based on an ordered set of publications in a reference set whereby the fraction of papers at or below the citation counts of a paper is used as a standardized value for the relative citation impact of the paper under study. For example, a percentile value of 90 means that the paper belongs to the 10% most-frequently cited papers which were published in the same field, as the same document type, and in the same publication year. The percentile values can be used for cross-field and cross-time-period comparisons.

While there are several methods to calculate percentiles (Bornmann, Leydesdorff, & Wang, 2013; Bornmann, Leydesdorff, & Wang, 2014), in this study the percentiles were calculated using Hazen's method (1914) because this method ensures that the mean percentile is 50 and symmetrically handles the tails of the distributions. When the normalized citation impacts are needed for more than a single paper, the percentile calculations are repeated for each paper using corresponding reference sets, respectively. If the paper is published in a journal with multiple WoS categories, a percentile is calculated for each category and the average of these percentile values is used (Bornmann, 2014).

Percentile classifications facilitate categories of the most frequently cited papers by country as follows: By fixing thresholds to percentiles of 90% and 99%, sets of papers by country are identified in the top 10% ($P_{top\ 10\%}$) and 1% ($P_{top\ 1\%}$) most frequently cited papers for 1990 to 2010 (and not later than 2010). The 2010 cut-off is made in order to have sufficient time for a three-year citation window which is needed to produce reliable citation impact scores (Wang, 2013).

**2.2 Software and statistical methods**

Stata is used for the statistical analyses (StataCorp., 2013). Pajek is used to create the co-authorship networks (http://pajek.imfm.si/doku.php; de Nooy, Mrvar, & Batagelj, 2011) applying the spring embedder of Kamada and Kawai (1989). Degree distribution is measured in the networks to view the pattern of connectivity, following de Nooy, et al. (2011, p. 81): The countries in the cluster are tightly connected "because each vertex has a particular minimum degree within the cluster … These clusters are called *k*-cores and *k* indicates the minimum degree of each vertex within the core" (p. 81).

# 3 Results

## 3.1 National representations within the most frequently cited papers

Table 1 lists the countries whose addresses appear in the data set in descending order of citation strength. Column 2 shows the number of records per country. Column 3 shows the number of papers calculated as falling in the top 10%. Column 4 shows the percentage of that country's articles appearing in the top 10%. For stochastic reasons, we can expect that each country will publish 10% of its papers in the top-10% segment of most frequently-cited papers, and similarly 1% in the 1% most frequently-cited papers, *ceteris paribus*. Columns 5 and 6 show the data for the top 1% of most frequently cited papers.

Table 1. The 30 countries worldwide with highest percentage of most frequently-cited papers (sorted in descending order by $PP_{top10\%}$). In order to include all BRICS countries, the list includes only those countries worldwide with more than 98,000 articles, reviews, and letters between 1990 and 2010 (with 98,635 papers, South Africa has the lowest number of papers in the list). The five and 15 best-performing countries are colored grey, as are the BRICS countries.

| *Country* | *Records* | $P_{top\,10\%}$ | $PP_{top10\%}$ | $P_{top\,1\%}$ | $PP_{top1\%}$ |
|---|---|---|---|---|---|
| Switzerland | 314,566 | 49,275 | 15.7 | 5,859 | 1.9 |
| Denmark | 170,960 | 25,022 | 14.6 | 2,832 | 1.7 |
| Netherlands | 445,353 | 64,667 | 14.5 | 7,060 | 1.6 |
| USA | 6,000,636 | 858,703 | 14.3 | 96,146 | 1.6 |
| Sweden | 324,221 | 41,792 | 12.9 | 4,327 | 1.3 |
| Belgium | 234,069 | 29,419 | 12.6 | 3,102 | 1.3 |
| UK | 1,613,321 | 201,588 | 12.5 | 20,855 | 1.3 |
| Norway | 119,297 | 14,312 | 12.0 | 1,493 | 1.3 |
| Canada | 837,922 | 100,307 | 12.0 | 10,474 | 1.2 |
| Finland | 154,506 | 18,247 | 11.8 | 1,837 | 1.2 |
| Australia | 517,442 | 58,612 | 11.3 | 5,854 | 1.1 |
| Germany | 1,425,655 | 159,250 | 11.2 | 15,738 | 1.1 |
| Austria | 160,209 | 17,785 | 11.1 | 1,919 | 1.2 |
| France | 1,052,832 | 112,965 | 10.7 | 10,971 | 1.0 |
| Israel | 208,203 | 22,266 | 10.7 | 2,180 | 1.0 |
| New Zealand | 100,876 | 10,361 | 10.3 | 1,026 | 1.0 |
| Italy | 732,577 | 74,378 | 10.2 | 7,150 | 1.0 |
| Spain | 539,510 | 50,797 | 9.4 | 4,526 | 0.8 |
| Greece | 125,522 | 10,134 | 8.1 | 913 | 0.7 |
| China | 979,740 | 75,537 | 7.7 | 6,827 | 0.7 |
| Japan | 1,428,823 | 109,249 | 7.6 | 9,371 | 0.7 |

| Taiwan | 249,817 | 18,612 | 7.5 | 1,332 | 0.5 |
| South Africa | 98,635 | 7,159 | 7.3 | 661 | 0.7 |
| Korea | 352,143 | 25,233 | 7.2 | 2,037 | 0.6 |
| Brazil | 285,423 | 16,025 | 5.6 | 1,309 | 0.5 |
| Mexico | 110,321 | 6,169 | 5.6 | 531 | 0.5 |
| Turkey | 193,031 | 10,100 | 5.2 | 793 | 0.4 |
| Poland | 235,507 | 12,042 | 5.1 | 1,170 | 0.5 |
| India | 462,315 | 22,320 | 4.8 | 1,530 | 0.3 |
| Russia | 489,879 | 15,887 | 3.2 | 1,413 | 0.3 |

The list of countries in Table 1 shows that, during the decades 1990-2010, Switzerland was the best-performing country with a proportion $P_{top\ 1\%}$ ($PP_{top\ 1\%}$) of 1.9% and $PP_{top\ 10\%}$ of 15.7%. Switzerland produced about twice more $P_{top\ 1\%}$ papers than expected, and about 1.6 times more $P_{top\ 10\%}$ papers. Switzerland is followed by 14 other countries with $PP_{top\ 1\%}$ and $PP_{top\ 10\%}$ above expected values: Denmark, the Netherlands, USA, Sweden, Belgium, UK, Norway, Canada, Finland, Australia, Germany, Austria, France, and Israel. This set of countries will be used in the network analysis below as the cohort for comparison with the five BRICS countries: how do the BRICS countries relate, in which segments, and in which years?

China is the best-performing BRICS country, with $PP_{top\ 1\%}$=0.7% and $PP_{top\ 10\%}$=7.7%. Although China is also among the ten countries with the highest aggregated publication output between 1990 and 2010 (the other nine countries are the USA, UK, Japan, Germany, France, Canada, Italy, Spain, and Australia), China's percentages of the most frequently-cited papers are below the expectations of 1% and 10%, respectively. Among the BRICS countries, China is followed in descending order by South Africa ($PP_{top\ 1\%}$=0.7% and $PP_{top\ 10\%}$=7.3%), Brazil ($PP_{top\ 1\%}$=0.5% and $PP_{top\ 10\%}$=5.6%), and India ($PP_{top\ 1\%}$=0.3% and $PP_{top\ 10\%}$=4.8%). Russia is at the bottom of the list with $PP_{top\ 1\%}$=0.3% and $PP_{top\ 10\%}$=3.2%, performing below expectation.

**3.2     Comparison of BRICS countries with the top five countries worldwide**

The top five countries from Table 1 are used as the comparison set for the BRICS. (Adding more countries would unnecessarily overload the figures without improving the calculations.)

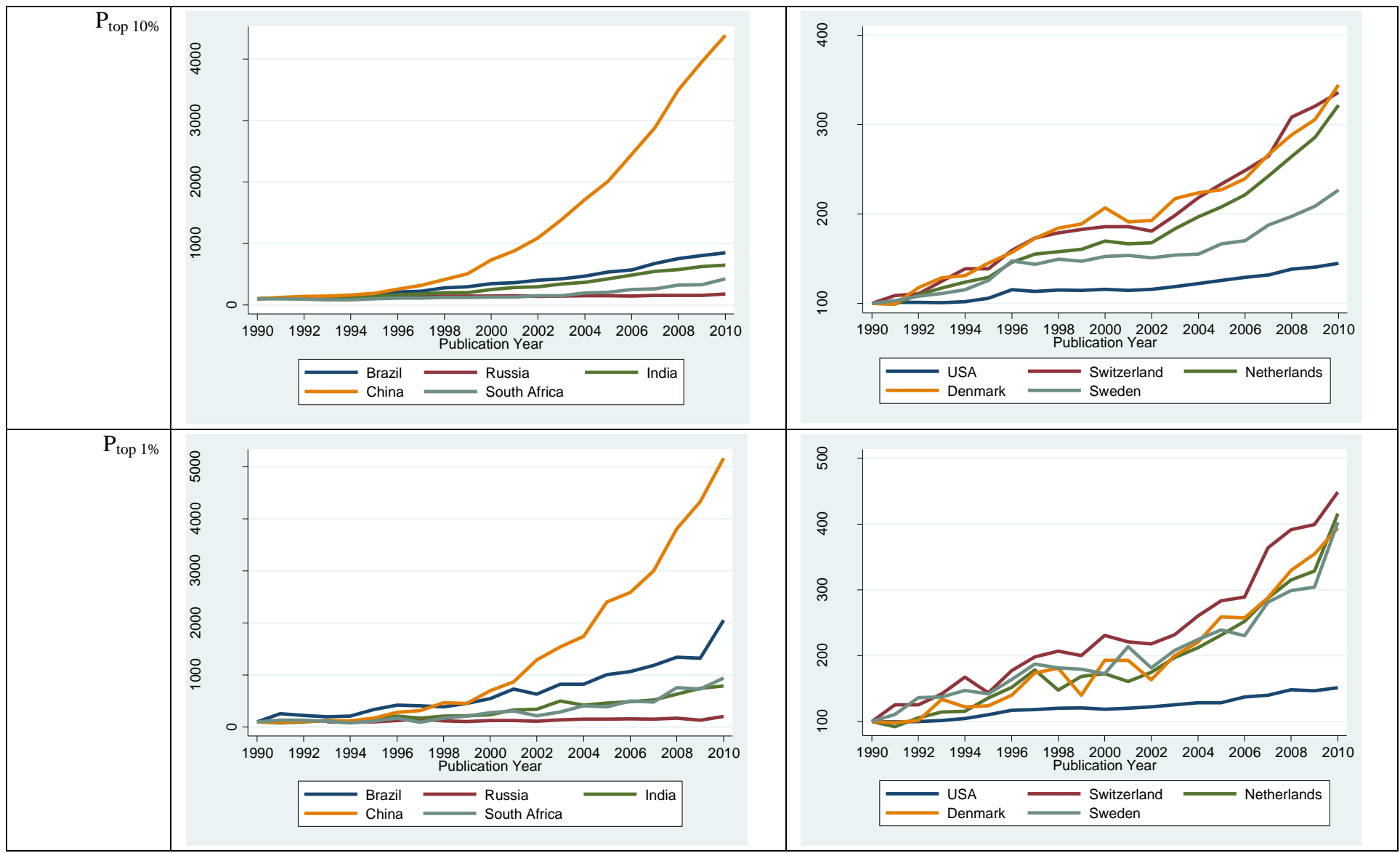

Figure 1. Number of $P_{top\ 10\%}$ and $P_{top\ 1\%}$ for the BRICS countries in comparison with the five best-performing countries worldwide relative to the year 1990. The number of publications in 1990 is the reference value for the publication numbers in the following years.

Figure 1 compares the numbers of $P_{top\ 10\%}$ and $P_{top\ 1\%}$ for the BRICS countries to the five top countries worldwide, setting 1990 as the base year. This is done to allow the calculation of the percentage change over time. For example, the USA has $P_{top\ 10\%}$=34,816 in 1990, and $P_{top\ 10\%}$=50,417 in 2010, resulting in a percentage of 145% for 2010 compared to 1990. Figure 1 shows that the best-performing BRICS countries (especially China, but also Brazil in terms of $P_{top\ 1\%}$) outperform all other comparison countries. Where the best-performing BRICS countries reach percentages of more than 500%, even the Switzerland, Denmark, the Netherlands, and the USA are definitely lower than 500%. Russia is the worst-performing BRICS country with only a slight growth. The USA shows the lowest increase in most frequently cited papers over the years. Although the USA dominates science in terms of sheer numbers and the USA produces a continuous stream of papers on a high level, the growth rate of the number of American papers is relatively low.

China shows exceptional increase in the numbers of the most frequently-cited papers: No other country (BRICS or otherwise) comes close to these extremely high growth rates. Similar results have been published elsewhere based on all records from literature databases (e.g., Adams, et al., 2013; Kumar & Asheulova, 2011). Russia falls below expectation with the lowest increase among the most frequently-cited papers (Kozak, Bornmann, & Leydesdorff, in preparation).

Figure 2 shows year-on-year annual growth rates for the two percentile rank classes, comparing the BRICS with other top-performing countries (Becketti, 2013). The following example explains how we calculated the growth rates: Researchers from Brazil, for example, published $P_{top\ 10\%}$=200 in 1990 and $P_{top\ 10\%}$=230 in 1991. Consequentially, the annual growth rate from 1990 to 1991 is 0.15 (=(230/200)-100). The annual growth rates calculated in this way significantly oscillate over the years, so we smoothed them by running medians of 3-year spans in combination with an outlier-resistant nonlinear smoothing technique (Velleman & Hoaglin, 1981). These smoothed growth rates are shown in Figure 2.

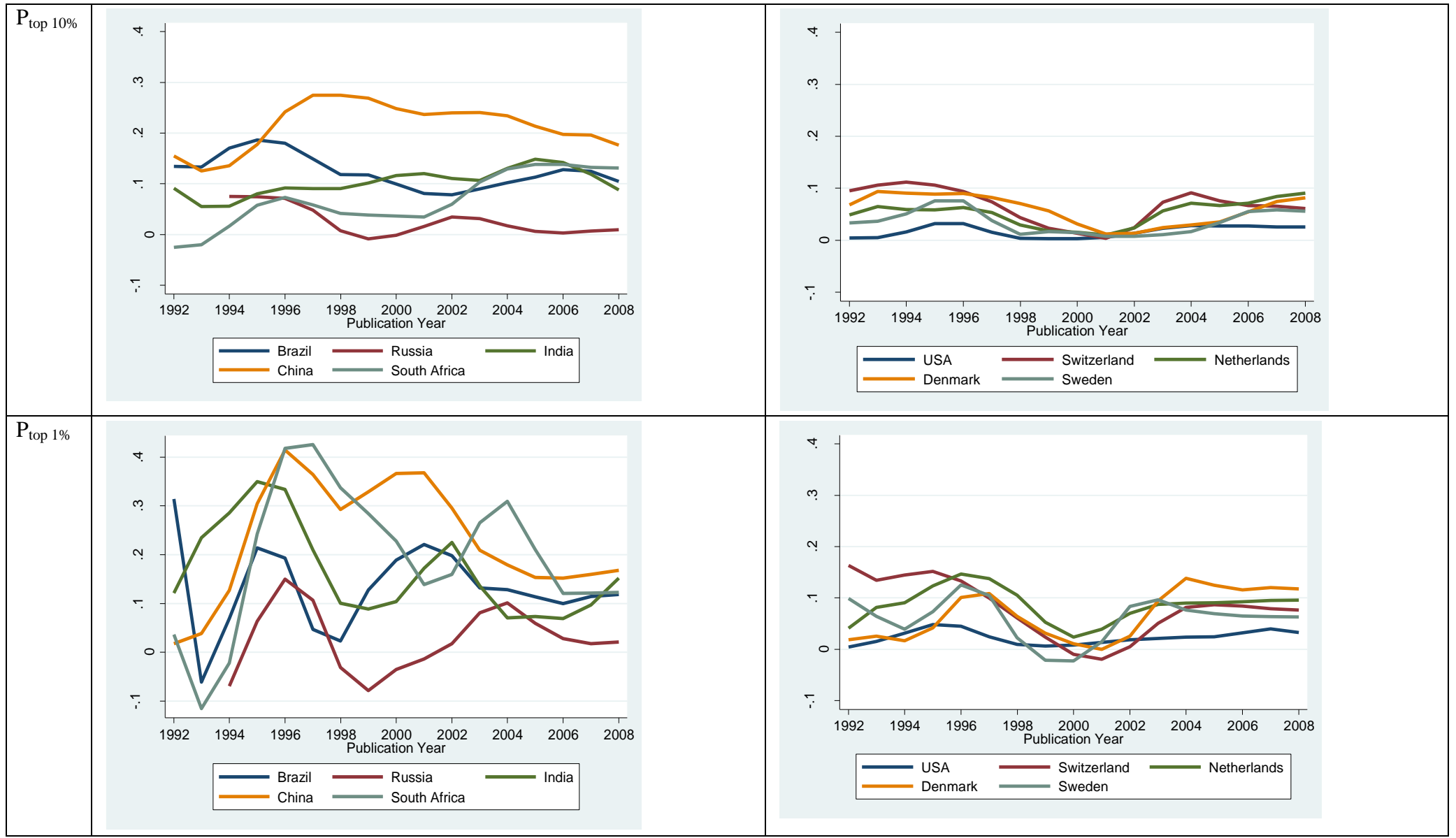

Figure 2. Annual growth rates of $P_{top\ 10\%}$ and $P_{top\ 1\%}$ for the BRICS countries compared to the five best-performing countries worldwide. The growth rates have been smoothed by Hanning and span-3 median smoothers

Since the smoothing of the growth rates needs previous and subsequent years, the publication years in this figure were restricted to the period from 1992 to 2008. In other words, the years 1990 and 1991 as well as 2009 and 2010 are omitted, because it is not possible to calculate running medians of 3-year spans in these cases.

Top-performing countries worldwide have growth rates between 0 and .1 as shown in Figure 2 with three remarkable deviations: The smaller countries (Switzerland, the Netherlands, and Denmark) have somewhat higher growth rates in terms of the $P_{top\ 1\%}$. The USA performs worst compared to the other countries. Around 2000, all best-performing countries show a significant decrease in growth rates with a fast rebound. A similar decrease was also visible in analyses based on all WoS publications (Bornmann & Mutz, in preparation). With the exception of Russia, the BRICS show higher annual growth rates than the best-performing countries worldwide. Using this index, the best-performing country among the BRICS countries is again China: Starting in the middle of the 1990s China has remarkably high growth rates, which decrease towards the most recent years, though. This decrease is also visible for the other BRICS countries. It seems that a period marked by huge growth rates waned in recent years with the BRICS countries tending to converge to the growth rates of the best-performing countries.

**3.3   Co-authorship networks of BRICS and the 15 top countries worldwide**

We follow a long line of literature acknowledging the importance of coauthorship as an indicator of collaboration (Katz & Martin, 1997; Persson, Glänzel, & Danell, 2004; Wagner & Leydesdorff, 2005). Figure 3 shows the co-authorship networks of the BRICS and the 15 top performing countries worldwide for the publication years 1995, 2000, 2005, and 2010 based on $P_{top\ 10\%}$ and $P_{top\ 1\%}$. Since we are interested in exploring the extent to which BRICS are integrated in the network of best-performing countries, we included all countries with both $PP_{top\ 10\%}$ and $PP_{top\ 1\%}$ above the expected values of 10% and 1% (shown in Table 1).

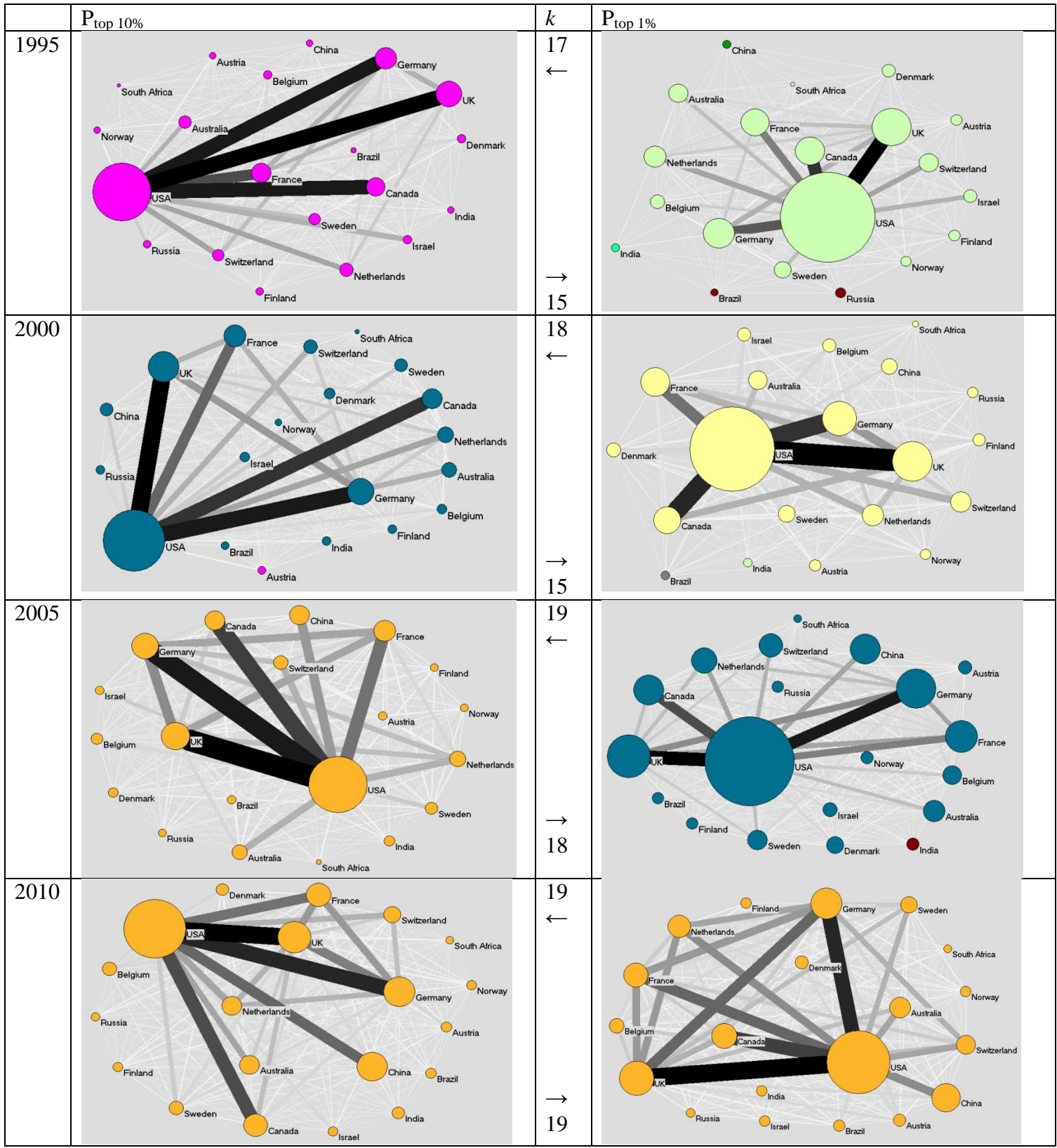

Figure 3. Co-authorship networks of the BRICS and the 15 best-performing countries worldwide for the publication years 1995, 2000, 2005, and 2010 based on $P_{top\ 10\%}$ and $P_{top\ 1\%}$. $k$ indicates the minimum degree of each vertex within the core of each network. For example, each vertex (light green) in the core set of the $P_{top\ 1\%}$ network for 1995 has a particular minimum degree within the cluster of $k$=15.

The size of the vertices in the figure reflects the numbers of $P_{top\ 10\%}$ and $P_{top\ 1\%}$ for each country, respectively. The sizes are not comparable across networks: Each network has been scaled differently. The thicker and darker the edges between two countries, the more frequently they are both named in a publication's address field.

The USA dominates the co-authorship relations in all years as shown in Figure 3. This is stable over time. In an historic shift, China becomes an important collaborating partner of the USA with remarkable growth seen in 2005 and 2010. The *k*-core measure shows that at the $P_{top\ 10\%}$ level all countries are connected to each other by one or two steps (with the exception of Austria in 2000 which may be a size anomaly) showing that the BRICS became part of the network of the best-performing countries. This picture changes if we look at the $P_{top\ 1\%}$ level. In 1995, only South Africa (not yet considered a "BRICS country") is part of the core set of the top-performing countries. Over the years, one BRICS country after another has been included in the core leading eventually to the inclusion of all BRICS countries in 2010.

## 4  Discussion

The governments of the BRICS countries boosted their investments in research and development to become part of the group of nations doing research at the highest level (UNESCO, 2010). The investigation presented here shows that the BRICS countries have cohorts of scientists producing papers in the domain of excellent research ($P_{top\ 10\%}$ and $P_{top\ 1\%}$). The five BRICS countries compete with the top-performing countries worldwide for elite status, and they cooperate with these leading countries in an elite network of communications.

This study depends upon the data available in the WoS. The figures presented in this study may be inflated by changes in the size of the database, although we do not know the extent to which this may be the case. WoS was significantly expanded in 2009 in order to enlarge the regional coverage (Testa, 2011), and also in response to competition from Scopus, which entered the market in 2004. Despite these relatively recent expansions, the developments seem to indicate growth, integration, and a return to the mean growth rates in more recent years for the BRICS countries. The dynamics of globalization may also have suffered from the crisis in the economies of the advanced nations since 2008 that may have an influence on exchange programs for post-docs, for example.

The data suggest that an exceedingly robust global science system has emerged, one that is open to new entrants from the BRICS countries, based upon merit. Communications among scientists appears to be growing considerably, moving some practitioners from countries—which did not participate in global science a generation ago—into an international communications network of recognition and exchange.

# Acknowledgements

The data used in this paper is from a bibliometrics database developed and maintained by the Max Planck Digital Library (MPDL, Munich) and derived from the Science Citation Index Expanded (SCI-E), Social Sciences Citation Index (SSCI), Arts and Humanities Citation Index (AHCI) prepared by Thomson Reuters (Scientific) Inc. (TR®), Philadelphia, Pennsylvania, USA: ©Copyright Thomson Reuters (Scientific) 2014. We would like to thank an anonymous reviewer for his/her valuable feedback to improve the paper.